\documentstyle[12pt,epsf]{article}

\input epsf.sty
\begin{document}

\begin{titlepage}

\title{
\begin{flushright}
{\small 
\hspace{1ex} hep-ph/0312030\\[-3ex]
KIAS--P03087, 
\hspace{1ex} VEC/PHYSICS/P/2/2003-2004 
} 
\end{flushright}
Supersymmetric threshold corrections to $\Delta m^2_{\odot}$}

\author{Biswajoy Brahmachari$^*$ and Eung Jin
Chun$^\dagger$\\[1ex]
$^*$ {\normalsize Department of Physics, Vidyasagar Evening College}\\
{\normalsize 39, Sankar Ghosh Lane, Kolkata 700006, India} \\[1ex]
$^\dagger$ {\normalsize Korea Institute for Advanced Study}\\
{\normalsize 207-43 Cheongryangri 2-dong, Dongdaemun-gu}\\
{\normalsize Seoul 130-722, Republic of Korea} \\}

\date{}
\maketitle

\abstract{
For nearly degenerate neutrinos, 
quantum corrections can modify the tree-level
masses via low energy supersymmetric threshold corrections comparable to
the solar oscillation mass scale. We numerically calculate corrections to
neutrino masses in minimal supergravity (mSugra) and Gauge Mediated
Supersymmetry Breaking (GMSB) scenarios and identify parameter spaces in
the high energy regime for which the solar neutrino mass splitting
becomes too large compared to the LMA solution.
We show that such considerations can give bounds on GMSB and
mSugra models which can be useful. On the contrary, if we start 
from degenerate mass eigenvalues at the tree level, these threshold
corrections being generation dependent, can also produce the required mass
splitting at solar scale for regions of parameter space.
}

\end{titlepage}

\section{Introduction}

Recent developments in neutrino experiments have provided fairly
significant information on the neutrino mass and mixing
parameters.  The new analysis of Super-Kamiokande collaboration
indicates that the atmospheric neutrino mass-squared difference
and  mixing angle satisfy $\Delta m^2_A = 1.3-3.0$ eV$^2$ and
$\sin^22\theta_A > 0.9$  \cite{ATM}.  A global analysis of all
solar neutrino data  yields $\Delta m^2_\odot = 7.1 ^{+1.2}
_{-0.6} \times 10^{-5}$ eV$^2$ and $\theta_\odot = 32.5 ^{+2.4}
_{-2.3}$ degrees including the KamLAND \cite{KamL} and SNO salt
results \cite{SOL}.  One of the important unknowns in the neutrino
sector is the structure of absolute mass scales which cannot be
determined by oscillation experiments. For this we can turn to
neutrino-less double beta decay experiments, cosmological bounds on
neutrino masses such as the WMAP bound and so on as explained next.

If neutrinos are almost degenerate (that is the mass splitting is
negligible compared to the masses), they could lead to an
observational signature in the future nuclear or
astrophysical/cosmological experiments. At present, there are
several upper limits on the absolute neutrino mass scale. Tritium
$\beta$ decay experiments put $m_\beta < 2.2$ eV \cite{Tbeta}, and
neutrino-less double beta decay experiments constrain the
effective Majorana  mass; $|m_{ee}| < 0.3-1.3$ eV depending on the
uncertainty in the nuclear matrix element \cite{Dbeta}.  The WMAP
collaboration has drawn the impressive limit on the sum of three
neutrino masses, $\sum_i m_i < 0.71$ eV, or equivalently, $m_i <
0.23$ for three degenerate neutrinos \cite{WMAP}. However, the
cosmological bounds are based on some assumptions and models,
depending on which one sets $\sum_i m_i < 1.1$ or $2.12$ eV
\cite{Hann}.

The nearly degenerate neutrino mass pattern is vulnerable to
quantum corrections.  Its stability has been studied extensively
in the context of the see-saw mechanism  where the renormalization
group evolution (RGE) \cite{RGE1,RGE2} can produce too large
corrections to keep the required mass degeneracy \cite{RGE3}.
Apart from the RGE effect, there can be another type of quantum
corrections, the low energy threshold effect.

For a sizable threshold corrections, one needs a large Yukawa
coupling effect or a large splitting between slepton masses in
supersymmetric theories.  The latter can arise in SO(10) models
with the top quark coupling effect on the RGE from the Planck
scale to the GUT scale \cite{chun-poko} or in a minimal
supersymmetric standard model (MSSM) with non-universal soft terms
\cite{cipv-c}. The general computation of the threshold
corrections in the Standard Model and in the MSSM has been made in
\cite{chan-waso}.
Note that the threshold corrections can arise independently of the
RGE effect in the seesaw mechanism and thus should be present in
any mechanism of generating the neutrino mass matrix \cite{chan-poko}.
This corrections to neutrino masses are generated by loop
corrections.

In this paper, we will consider the low energy threshold
corrections in the MSSM with minimal flavour violation, where the
flavour dependent structure arise only from the usual Yukawa
couplings and thus the supersymmetry breaking is taken to be
flavour blind. This is usually assumed in the MSSM to avoid the
dangerous supersymmetric flavour problems. Two popular scenarios of
such are the minimal supergravity (mSugra) model  and the gauge
mediated supersymmetry breaking (GMSB) models \cite{martin}. The
sources of  sizable threshold corrections are the tau Yukawa
coupling and the slepton mass splitting driven by it.  As a
consequence, we find  that the solar neutrino mass splitting can
arise solely through the threshold effect or constrains some
parameter space where $\tan\beta$ and the scalar and gaugino soft
masses are large.

Our consideration readily applies to low energy models of 
neutrino masses in which almost degenerate mass eigenvalues are
generated by some mechanism around the electroweak scale. 
Our results are independent of the form of neutrino
mass textures, while they depend on the pattern of
eigenvalues. Note that degenerate eigenvalues can
be obtained from many different mass textures.
Therefore, in this article we do not highlight
how a specific texture is obtained from a definite
flavor symmetry. If we invoke a specific flavor
symmetry our result will be less generally valid and
therefore weaker. We also find 
it is easier  to motivate degenerate neutrino mass spectrum
from an experimental point of view in view of
latest experimental results\cite{ATM,KamL,SOL}.

We give a few examples now to motivate our 
calculations event hough details of mass texture generation
is beyond the scope of the present article.

(a) The simplest possibility is to invoke a suitable Yukawa texture,
for instance, of the dimension-five operator ${f_{ij}\over M_R} L_i L_j 
\langle H \rangle \langle H \rangle$ in the see-saw mechanism with
a suitable low mass scale of right handed neutrino $\nu_R$ namely $M_R$.

(b) Alternatively, one could consider a low energy Higgs triplet as the
origin of neutrino mass generation \cite{degen3}, in which the resulting 
flavour violating signatures can be probed in the future experiments,
event hough our RGE analysis needs to be modified in the
presence of $SU(2)$ triplet scalars.
(c) More natural framework of generating a degenerate mass matrix 
is to impose certain flavor symmetries at low energy \cite{degen1},
sometimes realizing texture-zeros \cite{degen2}.  
Some models existing in literature can be non-supersymmetric.  However, 
it is rather straightforward to implement supersymmetry\footnote{
Note that supersymmetry is a space-time symmetry whereas
flavor symmetries are internal symmetries. Therefore supersymmetry
generators commute with generators of flavor symmetry under
consideration.} in such models\cite{general}. Therefore we
do not foresee serious problems if flavour scale is around
the electroweak scale, as long as the flavor symmetry
is either a global symmetry or a discrete symmetry.

If neutrino mass texture is generated at a sufficiently high scale, one
has to consider
as well the RGE effect which typically gives  a larger
correction than the threshold effect. For example, in the
usual see-saw mechanism, the RGE contribution is given
by $I_\tau \approx { h^2_\tau \over 16 \pi^2} \log {M_R \over M_{S}}$
where $M_S$ is the supersymmetry breaking scale.
For $\tan \beta \sim 50$, $M_S=M_Z$ and $M_R \sim 10^{10}$ GeV, we get
$I_\tau \sim 0.02$ which is an order of magnitude larger
than our threshold corrections as we will see later.  
The threshold corrections will also be useful if in some
case RGE effects cancel tree level mass generated at
high scale.  A typical example can be found in a class of models
for the radiative amplification of the mixing angles, in which case
the degeneracy of three masses should be stronger at the
electroweak scale than at a high scale \cite{radamp}.

\section{Radiative corrections to $\Delta m^2_\odot$}

Let us consider a tree-level neutrino mass matrix $M^0$ which  has
eigenvalues  $m^0_i$ and the mixing matrix $U^0$. 
In the tree-level
mass basis, the one-loop corrected mass matrix takes the form,
\begin{equation}
M_{ij} = m^0_i \delta_{ij} + {1\over2} I_{ij} (m^0_i + m^0_j).
\end{equation}
where $I_{ij}$ is the one-loop factor coming from wave-function
renormalization. It is often convenient to calculate radiative
corrections in  the flavour basis where the charged lepton masses
are diagonal.  Denoting the one-loop factor as $I_{\alpha\beta}$
in the flavour basis, we have the relation,
\begin{equation}
 I_{ij} = \sum_{\alpha, \beta} I_{\alpha\beta} U^0_{\alpha i}
U^0_{\beta j} \,.
\end{equation}
In the case of the minimal flavour violation in the MSSM, only the
diagonal components $I_{\alpha\alpha}$ are non-vanishing and they
satisfy $I_{ee}  = I_{\mu\mu} \neq I_{\tau\tau}$.   
The difference
between $I_{ee, \mu\mu}$ and $I_{\tau\tau}$  arises from the
sizable tau Yukawa coupling and the mass splitting between the 3rd
generation sleptons and the others. The equality $I_{ee}=I_{\mu\mu}$ is
deviated by the small electron and muon Yukawa couplings which 
can be safely ignored.   Then, one has
\begin{equation}
I_{ij}= I_{ee} \delta_{ij} + I_\tau \, U^0_{\tau i} U^0_{\tau j}
\end{equation}
where $I_\tau \equiv I_{\tau\tau} - I_{ee}$.  The overall factor
$I_{ee}$ can be dropped out and only $I_\tau$ can modify the tree
level result.

When the neutrino masses are nearly degenerate, $m^0_1\simeq m^0_2
\simeq m^0_3 \simeq m_\nu$, the quantum correction $I_\tau$ may
break up the degeneracy in a significant way.  The change in the
mass eigenvalues can be approximated by $m_i -m^0_i \simeq m_i^0
I_{ii} $ and thus we get
\begin{equation}
\Delta m^2_{ij} \simeq \Delta {m^0_{ij}}^2 + 2 m_\nu^2
(I_{ii}-I_{jj}) \,.
\end{equation}
Considering the mass-squared difference $\Delta m^2_{12}$ for the
solar neutrino oscillation, one finds that the loop correction can
produce the desired mass splitting if
\begin{equation}
I_\tau \simeq {\Delta m^2_\odot \over 2 \cos2\theta_{\odot}
s_{A}^2 m_\nu^2}
\end{equation}
where we have taken the standard parameterization of the mixing
matrix $U^0$; $U_{\tau 1} = s_{12} s_{23}$ and $U_{\tau 2} =
c_{12} s_{23}$ identifying $\theta_{12}=\theta_\odot$ and
$\theta_{23}=\theta_A$ to a good approximation of $\theta_{13}\ll
1 $.  Here, we remark that the above contribution  arises since
the solar neutrino mixing is not maximal \cite{RGE3}. Recall that
$\cos2\theta_\odot= 0.35-0.49$.  From the observed values of the
neutrino mass and mixing parameters mentioned in the Introduction,
one finds that the range of
\begin{equation}
I_\tau = (1.1 - 3.7) \times 10^{-3} \left( 0.3 \mbox{ eV} \over
m_\nu \right)^2
\end{equation}
is acceptable to generate solar neutrino mass-squared difference.
With the best-fit values,  we get $I_\tau \simeq 1.9 \times
10^{-3}$ for  $m_\nu=0.3 \mbox{ eV}$.

On the other hand, the threshold correction has to be
constrained so that $I_\tau < 3.7\times10^{-3} (0.3\mbox{
eV}/m_\nu)^2$, barring the cancellation between the tree-level and
one-loop contributions. This consideration will put some
constraint on the MSSM parameter space if $m_\nu > 0.3$ eV.
Therefore if we start from a high energy theory such as
mSugra or GMSB, and
evolve the supersymmetry breaking mass parameters from the
high energy theory to the low energy, we can identify
high energy parameter space for which the solar
mass splitting $m^2_\odot$ becomes too large compared
to the currently measured LMA region.

In this paper, we will assume no CP violation, that is, vanishing CP 
phases in neutrino mass matrix.  The RGE studies  showed that
both the mixing angles and mass eigenvalues
can be affected by the presence of phases \cite{rgephase}.
Similar phenomenon is expected to occur with threshold corrections, 
which we will leave for a future study.

\section{Supersymmetric threshold corrections with minimal
flavour violation}

In the MSSM with minimal flavour violation, the low energy
threshold correction is solely determine by the quantity
$I_\tau\equiv I_{\tau\tau}-I_{ee}$ defined in Eq.~(3). The
explicit formulae for the threshold corrections have been obtained
in Ref.~\cite{chan-waso}.  Adopting its result, we calculate
$I_\tau$ which consists of three contributions from the charged
Higgs boson, neutralinos and charginos as follows.

\smallskip

\underline{Charged-Higgs contribution}
\begin{equation}
16\pi^2 I_\tau(H^\pm) = {g^2\over2}{m_\tau^2 \over m_W^2}
      \left[ {1\over4}(1+\tan^2\beta)(-{1\over2}+\ln{m_{H^\pm}^2\over Q^2})
    +{1\over2}(1+{3\over2}\ln{m_{H^\pm}^2\over m_W^2}) \right]
\end{equation}
where $\tan\beta=v_2/v_1$.

\smallskip

\underline{Neutralino/sneutrino  contribution}
\begin{eqnarray}
16\pi^2 I_\tau(\chi^0(1)) &=&
     + {g^2+g^{\prime 2} \over 8} \sum_{j=1}^4 (s_W N_{1j} - c_W N_{2j})^2  \nonumber\\
   &&    [F(m_{\chi^0_j}^2, m^2_{\tilde{\nu}_\tau})
        -F(m_{\chi^0_j}^2, m^2_{\tilde{\nu}_e})] \\
16\pi^2 I_\tau(\chi^0(2)) &=& -{2\over v_2} \sqrt{g^2+g^{\prime
2}} \sum_{j=1}^4
      (s_W N_{1j} - c_W N_{2j}) N_{4j}m_{\chi^0_j} \nonumber\\
  &&     [G(m_{\chi^0_j}^2, m^2_{\tilde{\nu}_\tau})
        -G(m_{\chi^0_j}^2, m^2_{\tilde{\nu}_e})]
\end{eqnarray}
where $N_{ij}$ is the neutralino diagonalization matrix with the
flavour index $i$ corresponding to $\tilde{B}, \tilde{W}_3,
\tilde{H}^0_1$ and $\tilde{H}^0_2$ and the mass-eigenstate index
$j$ for the state $\chi^0_j$. The loop functions are defined by
$$F(x,y)=\ln{y \over Q^2} -{1\over2}+{x\over y-x}
    + {x^2\over (y-x)^2} \ln{x\over y} $$
$$  G(x,y)=\ln{y\over Q^2} -1 - {x\over y-x} \ln{x\over y} $$

\smallskip

\underline{Chargino/charged-lepton loop contribution}
\begin{eqnarray}
16\pi^2 I_\tau(\chi^\pm(1)) &=& + {g^2 \over 4} \sum_{j=1}^2
U_{1j}^2
     [F_{j+}- F(m_{\chi^\pm_j}^2,m_{\tilde{e}}^2)
          + {m_{LL}^2 - m_{RR}^2 \over
     m_{\tilde{\tau}_1}^2  - m_{\tilde{\tau}_2}^2} F_{j-} ] \nonumber\\
  &&  + {\sqrt{2}\over4} g \sum_{j=1}^2 U_{1j} U_{2j}
         {m_\tau^2 (A_\tau -\mu \tan\beta) \over
            v_1 (m_{\tilde{\tau}_1}^2  - m_{\tilde{\tau}_2}^2)}
     F_{j-}  \nonumber\\
  &&  + {1\over2} {m_\tau^2 \over v_1^2} \sum_{j=1}^2 U_{2j}^2
       [F_{j+} - {m_{LL}^2 - m_{RR}^2 \over
          m_{\tilde{\tau}_1}^2  - m_{\tilde{\tau}_2}^2} F_{j-} ]  \\
16\pi^2 I_\tau(\chi^\pm(2)) &=&
     -{\sqrt{2}\over v_2} g  \sum_{j=1}^2 U_{1j} V_{2j} m_{\chi^\pm_j}
     [G_{j+} - G(m_{\chi^\pm_j}^2,m_{\tilde{e}}^2)
     + {m_{LL}^2 - m_{RR}^2 \over
     m_{\tilde{\tau}_1}^2  - m_{\tilde{\tau}_2}^2} G_{j-} ] \nonumber\\
 && -{2\over v_2}  \sum_{j=1}^2  U_{2j} V_{2j} m_{\chi^\pm_j}
      {m_\tau^2 (A_\tau -\mu \tan\beta) \over
            v_1 (m_{\tilde{\tau}_1}^2  - m_{\tilde{\tau}_2}^2)}
      G_{j-}
\end{eqnarray}
where $m_{LL}$ and $m_{RR}$ denote the left-handed and
right-handed stau masses whose mass eigenvalues are denoted by
$m_{\tilde{\tau}_{1,2}}$, $m_{\tilde{e}}$ denotes the left-handed
selectron (or smuon) mass, $U_{ij}$ and $V_{kj}$ are the chargino
diagonalization matrices  with the index $i=1,2$ for the flavour
states $\tilde{W}^-$, $\tilde{H}^-_1$, $k=1,2$ for $\tilde{W}^+$,
$\tilde{H}^+_2$, and $j=1,2$ for the mass eigenstate $\chi^\pm_j$.
Two loop functions are defined by
$$F_{j\pm}= {1\over2} [F(m_{\chi^\pm_j}^2,m_{\tilde{\tau}_1}^2)
         \pm F(m_{\chi^\pm_j}^2,m_{\tilde{\tau}_2}^2)], $$
$$G_{j\pm}= {1\over2} [G(m_{\chi^\pm_j}^2,m_{\tilde{\tau}_1}^2)
         \pm G(m_{\chi^\pm_j}^2,m_{\tilde{\tau}_2}^2)]. $$

\bigskip

Summing all the contributions, we get the total low-energy
threshold  correction to the neutrino mass matrix defined at the
scale $Q=M_Z$:
$$
I_\tau= I_\tau(H^\pm) + I_\tau(\chi^0(1)) + I_\tau(\chi^0(2))
+ I_\tau(\chi^\pm(1)) + I_\tau(\chi^\pm(2)).
$$
In the next section, we will analyze $I_\tau$ in models with
minimal gravity-mediated and gauge-mediated supersymmetry
breaking.

\section{Results in mSugra and GMSB models}

\begin{figure}
\begin{center}
\epsfysize=10cm \epsfxsize=10cm \epsffile{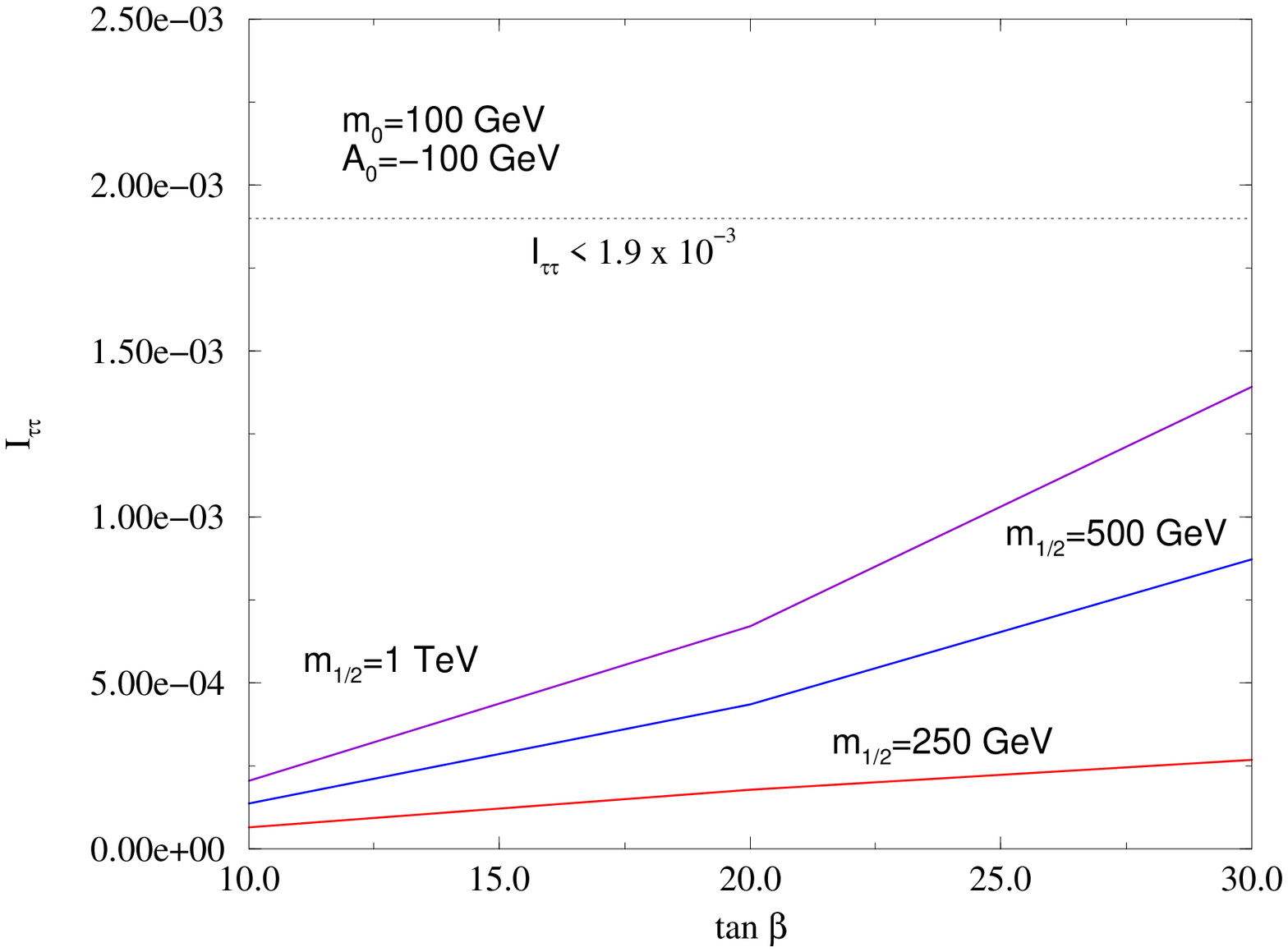} \caption{
Total amount of $I_{\tau}$ is plotted as a function of $\tan
\beta$. We have chosen three representative values of
$m_{1/2}=250,500,1000$ GeVs respectively. We have restricted
ourselves up to $\tan \beta=30$ because tachionic modes appear for
larger $\tan \beta$. For such small values of $m_0,m_{1/2}$, total
$I_{\tau}$ remains within acceptable limits } \label{fig1}
\end{center}
\end{figure}

\begin{figure}
\begin{center}
\epsfysize=10cm \epsfxsize=10cm \epsffile{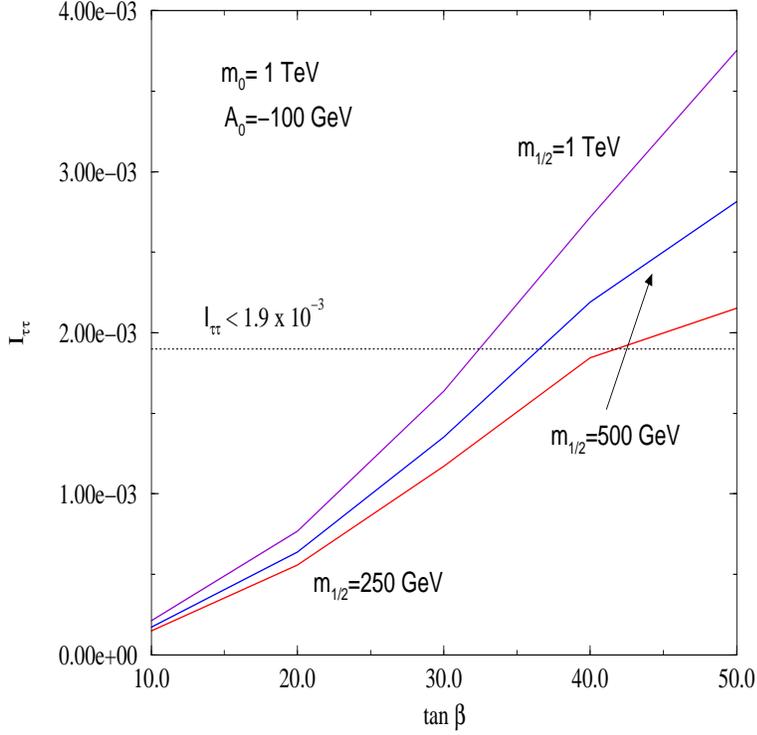} \caption{For
large $\tan \beta$ threshold corrections can produce $\Delta
m^2_\odot$ to fit LMA solution and put bounds on soft parameter space,
in particular for $\tan\beta > 50$.  The dotted line corresponds to  the 
best fit value of $\Delta m^2_\odot$ for the approximately degenerate
neutrino mass $m_\nu = 0.3$ eV.
} \label{fig2}
\end{center}
\end{figure}

Given the tree-level neutrino mass matrix $M^0$ with almost
degenerate eigenvalues at the weak scale, the threshold correction
derived in the above section can produce a significant change in
the neutrino mass splitting.  As one can see from Eqs.~(6-10), the
low energy threshold effect arises due to the flavour violation in
the Yukawa and  slepton sectors.  The latter is driven also by the
Yukawa coupling effect in the MSSM with minimal flavour violation,
we expect to have a sizable correction for large $\tan\beta$ and
large soft scalar masses $A_0$. In GMSB also large $\tan \beta$
region gives larger contribution than small $\tan \beta$ region.
However the overall corrections induced in the GMSB scenario is generally
smaller than the overall correction in mSugra scenario. This is
mainly because the splitting among soft masses in GMSB is relatively
smaller than  those of the mSugra case. For low $\tan\beta$ charged Higgs
dominates in the mSugra case as  can be seen  from Tables \ref{table1}
and \ref{table2}, whereas for large $\tan\beta$, typically charged
Higgs and chargino contributions are important for large $A_0$.
In mSugra, there is a large parameter space where the desired solar 
neutrino mass splitting can be generated.  However, 
for large $m_0$, $m_{1/2}$ and $\tan\beta$
solar splitting can be overshot and thus bounds on high energy
parameter space can also be obtained. This is displayed in Figures
\ref{fig1} and \ref{fig2}.  Let us note that the figures are 
generated by calculating some specific points connected by lines.
We also see
$A_0$ dependence of the result is very mild. Results also do not depend
appreciably on the sign of $A_0$. For the soft masses less than 1 TeV and
$\tan\beta \leq 50$, we find  $I_\tau < 4\times10^{-3}$, which is
marginally compatible with the limit $I_\tau < 3.7\times 10^{-3} (0.3
\mbox{ eV}/m_\nu)^2$. Stronger  bounds can be put for $m_\nu > 0.3$ eV.
In GMSB we typically have much small effect. Therefore
GMSB parameter space is generally compatible with the solar neutrino data
in the sense that chances of generating the solar splitting  
$\Delta m^2_\odot$ is much smaller in the GMSB case than
the mSugra case. These results are given in Table \ref{table3}.

In doing these calculations we have used SOFTSUSY program \cite{allan}
to calculate the low energy supersymmetry breaking soft parameters in
mSugra as well as GMSB scenarios of supersymmetry breaking.

\section{Conclusion}

If neutrino masses are almost degenerate, quantum corrections can
give rise to a significant effect on the neutrino mass and mixing
parameters.  One of important radiative corrections is the low
energy threshold effect which has to be added to the tree-level
mass matrix defined at the weak scale. In this paper, we have
considered such threshold corrections in the context of the
minimal supergravity and and gauge-mediated supersymmetry breaking
models where the lepton flavour violation arises only through the
usual Yukawa coupling effect.  At low energy, there are two
sources of threshold corrections; the tau Yukawa coupling and the
slepton mass splitting driven by it.  In mSugra models,  these two
effects become important to determine the solar neutrino mass
splitting when both the scalar and gaugino soft masses and
$\tan\beta$ are large.  As a consequence, the threshold correction
can provide a radiative origin of the solar neutrino mass
splitting or some constraints on the mSugra parameter space if the
overall neutrino mass scale is observed near the current
cosmological limit;   $m_\nu \sim 0.3$ eV. However we must keep in
mind that these numerical bounds can potentially be much stronger
if $m_\nu > 0.3$ eV. The effect turns out to be suppressed in the
GMSB models for typical ranges of parameter spaces at the high
energy scale.

{\bf Acknowledgments}: B.B. would like to thank Korea Institute
for Advanced Study for very kind hospitality and financial support
for one month. B.B. and E.J.C. would like to thank the organizers
of WHEPP-7 workshop where this work was initiated. We would also
like to thank Prof. Asim K. Ray for some discussions on neutrino
textures.

\newpage

\begin{table}
\begin{tabular}{|c|c|c|c|c|c|c|c|c|} \hline
CASE  & \multicolumn{3}{c}{mSugra parameters}& & $I_\tau(H^\pm)$ &
$I_\tau(\chi^0)$ & $I_\tau(\chi^\pm)$
& Total=$I_{\tau}$\\
\cline{2-5}
 & $m_0$ &$m_{1/2}$ & $A_0$ & $\tan \beta$ & &&& \\
\hline
1 & 100 & 250 & -100 & 10 & 6.0 $\times 10^{-5}$ & -2.7 $\times 10^{-7}$
& 4.1 $\times 10^{-6}$ & 6.38 $\times 10^{-5}$\\
2 & 100 & 250 & -100 & 20 & 2.2 $\times 10^{-4}$ & -1.8 $\times 10^{-5}$
& -2.4 $\times 10^{-5}$ & 1.78 $\times 10^{-4}$\\
3 & 100 & 250 & -100 & 30 & 4.5 $\times 10^{-4}$ & 7.1 $\times 10^{-4}$
& -1.2 $\times 10^{-4}$ & 1.04 $\times 10^{-3}$\\
\hline
\end{tabular}
\caption{This table shows the break-up of $I_{\tau}$ for typical
values of mSugra parameters and displays individual contributions
from individual $H^\pm,\chi^0$ and $\chi^\pm$ loops. For $m_0 \sim
100$ GeV tachionic modes appear for large $\tan \beta \sim 40$.}
\label{table1}
\end{table}

\newpage

\begin{table}
\begin{tabular}{|c|c|c|c|c|c|c|c|c|} \hline
CASE  & \multicolumn{3}{c}{mSugra parameters}& & $I_\tau(H^\pm)$ &
$I_\tau(\chi^0)$ & $I_\tau(\chi^\pm)$
& Total=$I_{\tau}$\\
\cline{2-5}
 & $m_0$ &$m_{1/2}$ & $A_0$ & $\tan \beta$ & &&& \\
\hline
1 & 1000 & 1000 & -100 & 30 & 1.1 $\times 10^{-3}$ & 3.3 $\times 10^{-5}$
& 5.4 $\times 10^{-4}$ & 1.67 $\times 10^{-3}$\\
2 & 1000 & 1000 & -100 & 40 & 1.8 $\times 10^{-3}$ & 6.5 $\times 10^{-5}$
& 8.9 $\times 10^{-4}$ & 2.75 $\times 10^{-3}$\\
3 & 1000 & 1000 & -100 & 50 & 2.4 $\times 10^{-3}$ & 1.1 $\times 10^{-4}$
& 1.2 $\times 10^{-3}$ & 3.71 $\times 10^{-3}$\\
\hline
\end{tabular}
\caption{This table shows the break-up of $I_{\tau}$ for
values of mSugra parameters $m_0=1000,m_{1/2}=1000,A_0=-100$ and
displays individual contributions from individual $H^\pm,\chi^0$
and $\chi^\pm$ loops.
} \label{table2}
\end{table}

\newpage

\begin{table}
\begin{tabular}{|c|c|c|c|c|c|c|c|c|} \hline
CASE  & \multicolumn{3}{c}{GMSB parameters}& & $I_\tau(H^\pm)$ &
$I_\tau(\chi^0)$ & $I_\tau(\chi^\pm)$
& Total=$I_{\tau}$\\
\cline{2-5}
 & $N_5$ &$M_{mess}$ & $\Lambda$ & $\tan \beta$ & &&& \\
\hline
1 & 1 & 1.61 $\times 10^{5}$
& 7.6 $\times 10^{4}$& 10
& 6.4 $\times 10^{-5}$ & -9.5 $\times 10^{-7}$
& 2.7 $\times 10^{-5}$ & 9.0 $\times 10^{-5}$\\
2 & 1 & 1.61 $\times 10^{5}$
& 7.6 $\times 10^{4}$ & 20
& 2.4 $\times 10^{-4}$ & -3.6 $\times 10^{-6}$
& 5.7 $\times 10^{-5}$ & 2.9 $\times 10^{-4}$\\
3 & 1 & 1.61 $\times 10^{5}$
& 7.6 $\times 10^{4}$ & 30
& 5.0 $\times 10^{-4}$ & -8.0 $\times 10^{-6}$
& 9.8 $\times 10^{-5}$ & 5.9 $\times 10^{-4}$\\
4 & 1 & 1.61 $\times 10^{5}$
& 7.6 $\times 10^{4}$ & 40
& 8.9 $\times 10^{-4}$ & -8.0 $\times 10^{-6}$
& 2.0 $\times 10^{-4}$ & 1.0 $\times 10^{-3}$\\
5 & 1 & 1.61 $\times 10^{5}$
& 7.6 $\times 10^{4}$ & 50
& 1.0 $\times 10^{-3}$ & -2.0 $\times 10^{-5}$
& 8.0 $\times 10^{-5}$ & 1.1 $\times 10^{-3}$\\
\hline
\end{tabular}
\caption{ GMSB models produce $\Delta m^2_\odot$ which are
fully consistent with the LMA region. $I_{\tau}$
and individual contributions are displayed in a typical GMSB model
with the messenger sector at around $10^5$ GeV or so. We see that
even at $\tan \beta=50$ the total $I_{\tau \tau}$ is well below
the acceptable limits.}
\label{table3}
\end{table}

\end{document}